\documentclass[aps,prl,twocolumn,showpacs,superscriptaddress,floatfix]{revtex4-1} 
\usepackage{amsmath,amssymb}
\usepackage{hyperref}
\usepackage{graphics,psfrag}
\usepackage{graphicx,psfrag}
\usepackage{epsfig}
\usepackage{longtable}
\usepackage{tabularx}
\usepackage{booktabs}
\usepackage[utf8x]{inputenc}
\usepackage[USenglish]{babel}
\usepackage{multirow}
\usepackage{rotating}
\usepackage{hhline}
\usepackage{xcolor}
\usepackage{ulem}
\usepackage[hang]{subfigure}
\usepackage{todonotes}
\usepackage{ulem}
\usepackage{nicefrac}
\usepackage{todonotes}

\begin{document}

\title{Charge storage in oxygen deficient phases of TiO$_2$: defect Physics without defects.}

\author{A. C. M. Padilha}
\email{antonio.padilha@ufabc.edu.br}
\affiliation{Centro de Ci\^encias Naturais e Humanas, Universidade Federal do ABC, Santo Andr\'e, SP, Brazil}
\affiliation{Department of Physics, Yokohama National University, Yokohama, Japan}

\author{H. Raebiger}
\affiliation{Centro de Ci\^encias Naturais e Humanas, Universidade Federal do ABC, Santo Andr\'e, SP, Brazil}
\affiliation{Department of Physics, Yokohama National University, Yokohama, Japan}

\author{A. R. Rocha}
\affiliation{Instituto de F\'isica Te\'orica, Universidade Estadual Paulista, S\~ao Paulo, SP, Brazil}

\author{G. M. Dalpian}
\email{gustavo.dalpian@ufabc.edu.br}
\affiliation{Centro de Ci\^encias Naturais e Humanas, Universidade Federal do ABC, Santo Andr\'e, SP, Brazil}

\date{\today}


\begin{abstract}
Defects in semiconductors can exhibit multiple charge states, which can be used for charge storage applications. Here we consider such charge storage in a series of oxygen deficient phases of TiO$_2$, known as Magnéli phases. These Ti$_n$O$_{2n-1}$ Magnéli phases present well-defined crystalline structures, \textit{i.\ e.}, their deviation from stoichiometry is accommodated by changes in space group as opposed to point defects. We show that these phases exhibit intermediate bands with the same electronic quadruple donor transitions akin to interstitial Ti defect levels in TiO$_2$-rutile. Thus, the Magnéli phases behave as if they contained a very large \textit{pseudo-defect} density: $\nicefrac{1}{2}$ per formula unit Ti$_n$O$_{2n-1}$. Depending on the Fermi Energy the whole material will become charged. These crystals are natural charge storage materials with a storage capacity that rivals the best known supercapacitors.
\end{abstract}


\maketitle

As our energy requirements grow, and alternative energy sources become an integral part of most countries' energy matrices, energy carriers, in particular charge storage systems play an ever increasing role. Li-ion batteries have played the major role in energy storage up to now \cite{Goodenough2013}, but new systems termed supercapacitors \cite{Yu2015} have emerged and are becoming more popular. In this case, a number of materials - mainly metal oxide thin films - provide charge storage due to the presence of defects inside its porous structure\cite{Sugimoto2003,Toupin2004,Simon2008,Young2015}. 

Titanium oxide is an important wide band gap semiconductor material for applications in photocatalysis \cite{Linsebigler1995,DiValentin2006,Kruger2012}, energy storage \cite{Zhang2015,Oh2014}, and memory devices \cite{Wilk2001,S_Williams_nature_2008,Kwon2010}. The stoichiometric phases of this material, TiO$_2$-rutile, -anatase, and -brookite are known to present unintentional n-type doping at ambient conditions, owing in part to intrinsic point defects related to oxygen deficiency: oxygen vacancy ($V_{\text{O}}$) and titanium interstitial ($\text{Ti}_i$) \cite{Lee2012,Janotti2010}. Further increase in oxygen deficiency leads to the formation of shear planes and consequently a phase transition to the so-called Magnéli phases \cite{Szot2011}.

\begin{figure}[h]
   \includegraphics[width=0.9\columnwidth]{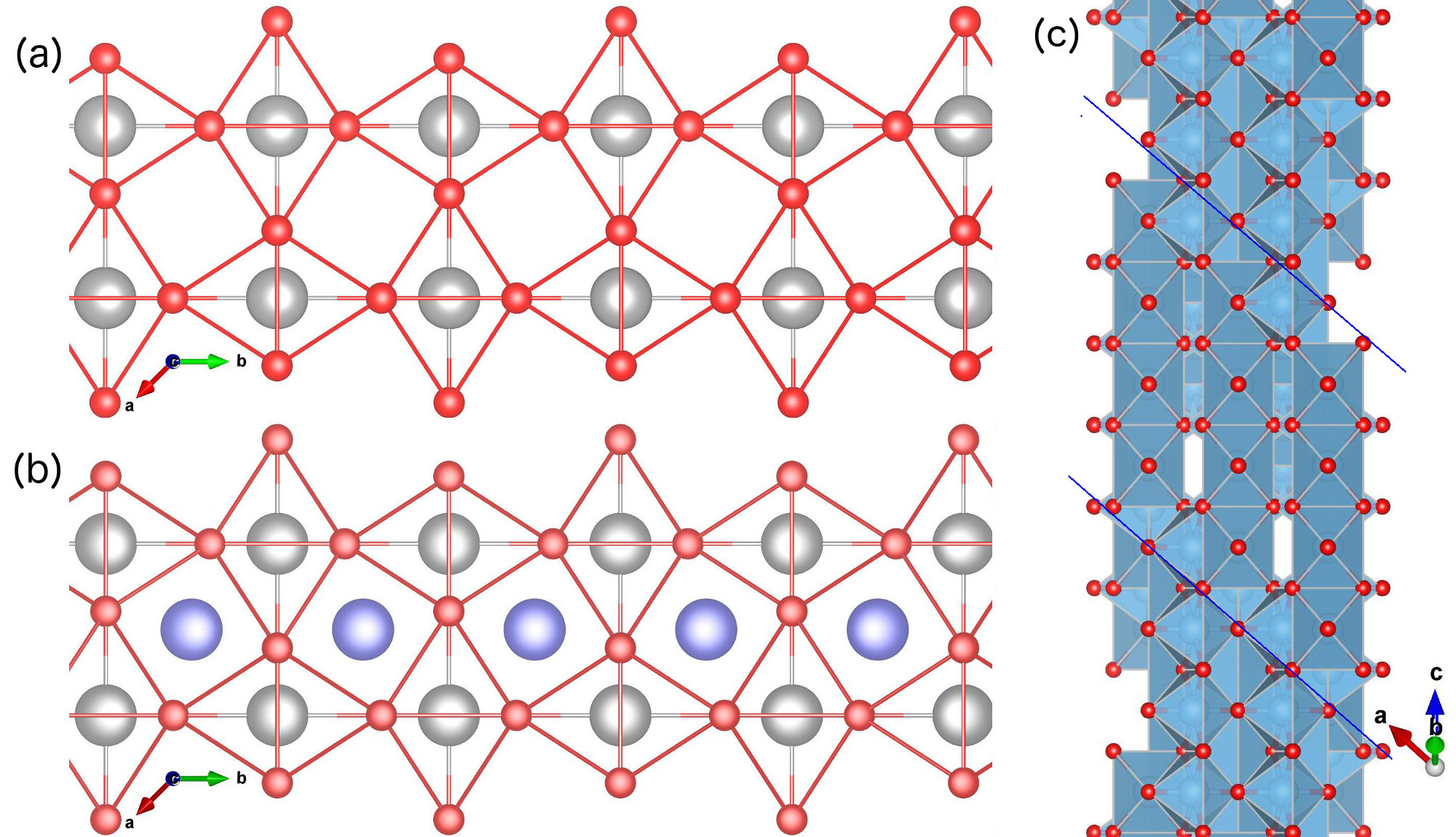}
   \caption{(color online) (a) TiO$_2$-rutile structure, (b) and (c) Ti$_4$O$_7$ Magnéli phase structures. The oxygen sub net is depicted in red. In (a) and (b) Ti atoms are either gray spheres when in standard rutile sites and blue spheres when in shifted positions while oxygen atoms are red spheres. In (c) the blue lines enclosure the four-units rutile-like chains along the $c$ direction, bounded by corundum-like planes restricted to the $(001)$ planes.}
   \label{fig:structs}
 \end{figure}

A few reports on the thermochemistry \cite{Liborio2008,Harada2010}, electrical properties \cite{Bartholomew1969}, and electronic structure \cite{Liborio2009,Weissmann2011,Leonov2006,Padilha2014,Padilha2015} of such systems are available, but the possibility of electronically charging them remains uncharted territory. Such charging becomes relevant as these materials are used as the active media of memristor devices \cite{Szot2011,Pan2014} or storage applications \cite{Zhang2015,Oh2014}, and in those cases, the exchange of electrons with a reservoir must be taken into account. 

In this letter we study the stability and electronic structure of Ti$_4$O$_7$ and Ti$_5$O$_9$ Magnéli phases, as well as of the Ti$_2$O$_3$ corundum phase, while in contact with a reservoir of electrons. We show that these TiO phases present a series of properties akin to $\text{Ti}_i$-containing TiO$_2$-rutile, such as mid gap states and charge state transitions. We show that the intermediate band typical for the Magnéli phases can donate electrons to an electron reservoir, leading to a new electronic phase that resembles charged defects in a semiconductor, even though they contain no crystallographic defects. The combination of such properties is shown to enable charge storage in these systems in such an efficient way that they can rival the best supercapacitors to date \cite{Yu2015}. 

The Magnéli phases have the general oxygen-deficient chemical formula Ti$_n$O$_{2n-1}$ ($n > 4$). In general, for $n > 37$ the crystal structure is still TiO$_2$-rutile, containing point defects or Wadsley defects. Further removal of oxygen (a decrease in $n$) leads to the reorganization of the crystal into these new crystallographic phases \cite{Bursill1969,Szot2011}. These phases can be described as being composed of rutile-like chains (edge- and corner-sharing arrangement) of $n$ TiO$_6$ octahedra units along the $c$ axis bounded by a corundum structure (\textit{i.\ e.} Ti$_2$O$_3$, composed of face-sharing TiO$_6$ octahedra) \cite{Marezio1971,Marezio1973,LePage1984}. From this point of view, these phases can be interpreted as an ordered combination of TiO$_2$-rutile and Ti$_2$O$_3$-corundum parts. The corundum-like boundaries of the rutile-like region of the Magnéli phases are usually referred to as shear planes. A model structure of these oxygen deficient phases can be obtained from rutile via a shear operation $(121)\frac{1}{2}[0\bar{1}1]$ \cite{Wood1981,Andersson1960,Padilha2015,Harada2010}. This operation can be understood as successive displacements of the atoms in the rutile crystal. All atoms above each $(121)$ plane shifted $n$ times along the $c$ vector from the origin are in turn dislocated in the $[0\bar{1}1]$ direction of the rutile structure. This direction coincides with a lattice vector of the oxygen subnet--\textit{i.\ e.}, the vector $[0\bar{1}1]$ connects two oxygen atoms in the rutile crystal--, thus it maps the dislocated atoms of that species into atoms of the same species, and finally leaves the lattice positions for oxygen atoms unchanged. Figure \ref{fig:structs} shows the structures of TiO$_2$-rutile and Ti$_4$O$_7$. From the perspective of (a) and (b) one can see that the oxygen deficiency of such compounds is better described by extra Ti atoms occupying interstitial positions of the TiO$_2$ matrix, rather than by oxygen atoms missing at specific lattice sites, \textit{i.\ e.}, oxygen vacancies.
\begin{figure}[h]
   \includegraphics[width=0.9\columnwidth]{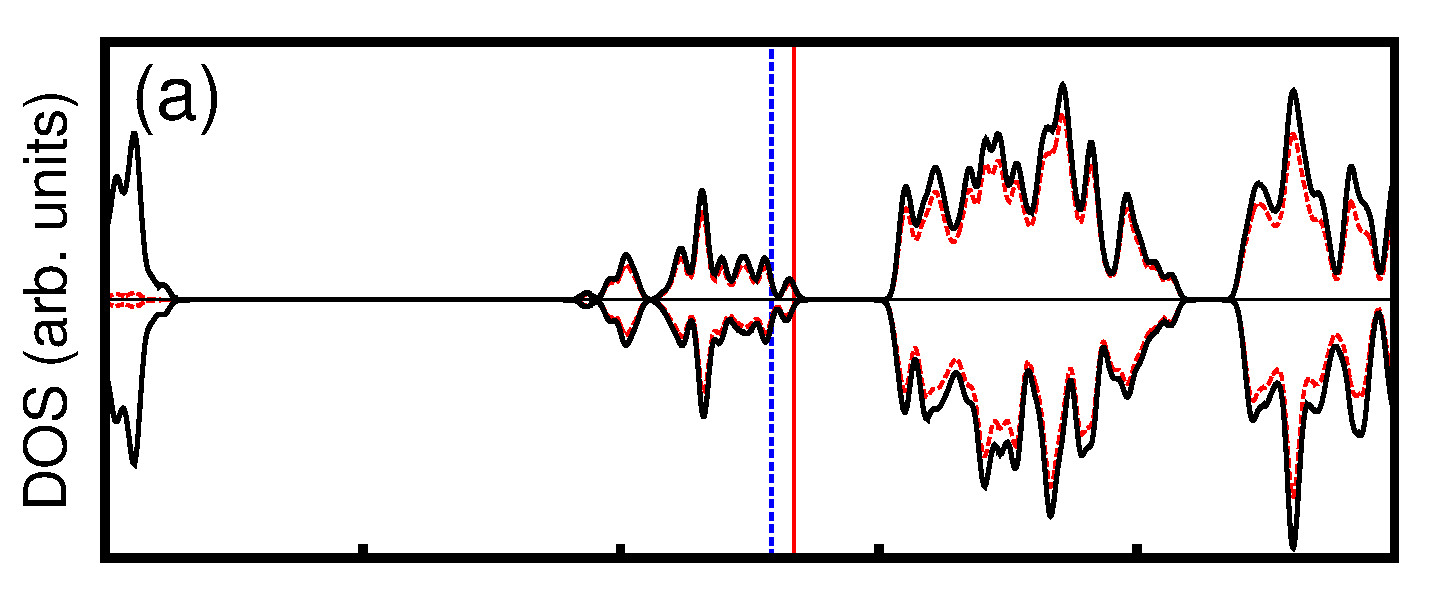}
   \includegraphics[width=0.9\columnwidth]{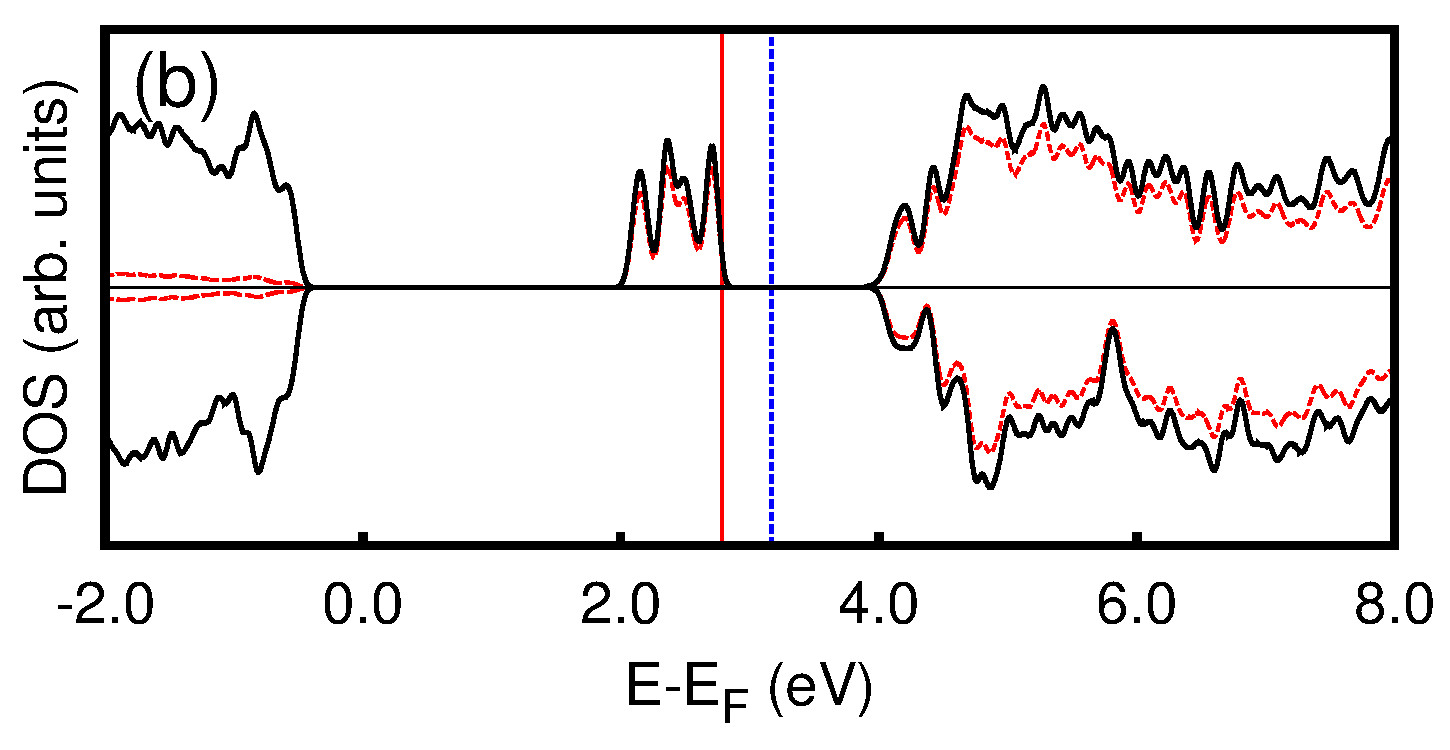}
   \caption{(color online) Projected density of states for (a) Ti$_2$O$_3$ and (b) Ti$_4$O$_7$. The spin components are distinguished by the upper and lower panels on each graph. The black full line is the total DOS and the red dashed line represents Ti(d) contribution. Energies are referenced from the last occupied level of the host material (TiO$_2$) by core-level (Ti 1s) shifts. The full vertical red line indicates the most energetic occupied level of each compound, while the vertical dashed blue line indicates the TiO$_2$ CMB.}
   \label{fig:dos-single}
 \end{figure}
 
\begin{figure}[t]
  \begin{center}
   \includegraphics[width=0.95\columnwidth]{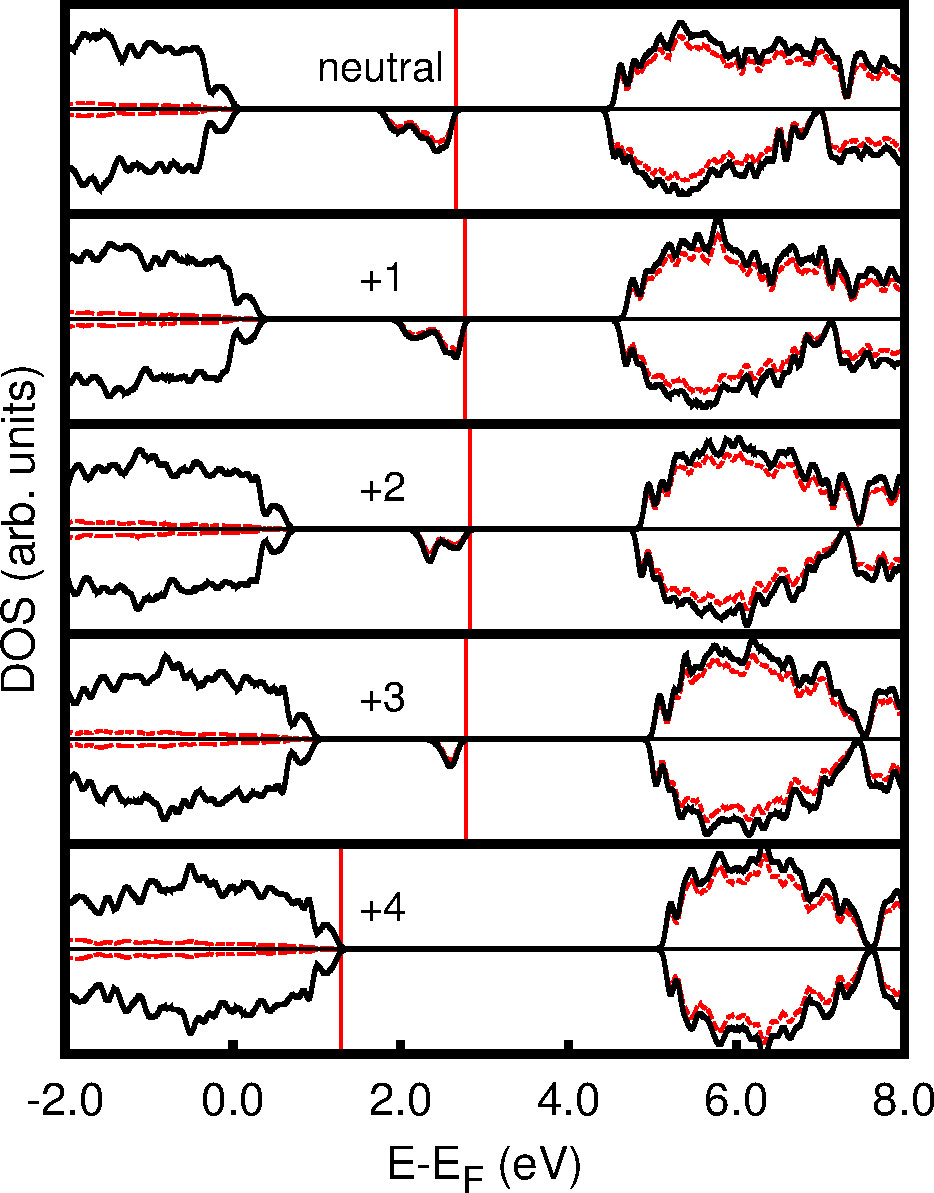}
   \caption{(color online) Projected Density of States (PDOS) for all charge states of Ti$_5$O$_9$. The full black line is the total DOS and the red dashed line represents Ti(d) contribution to the DOS. The two spin components are represented by positive and negative values along vertical axis. The vertical full red line indicates the last occupied electronic level, and the vertical dashed blue line indicates the TiO$_2$ CBM. Energies are referenced from the last occupied level of the host material (TiO$_2$-rutile).}
   \label{fig:dos-ti5o9} 
  \end{center}
 \end{figure}

From the electronic point of view, these oxygen-deficient Ti$_n$O$_{2n-1}$ phases present an \textit{intermediate band} \cite{Padilha2014,Liborio2009,Weissmann2011} slightly below the conduction band minimum (CBM). This is shown by the projected density of states (PDOS) given for Ti$_2$O$_3$ and Ti$_4$O$_7$ in Figure \ref{fig:dos-single}, and for Ti$_5$O$_9$ in Figure \ref{fig:dos-ti5o9} (upper panel). These DOS show striking resemblance to those observed for isolated defects in TiO$_2$\cite{Lee2012,Janotti2010}, and thus, we describe these states to be due to the presence of \textit{pseudo-defects} inside the Mangéli phases. As these phases present a high concentration of such \textit{pseudo-defects}, one can think of this intermediate band as the spatially-extended generalization of point defects. Importantly, this \textit{pseudo-defect} band  lies close to the TiO$_2$-rutile CBM, indicating that its occupation can be tuned by the use of appropriate leads, leading to charging of the material. We investigate this charging process by electronic structure calculations of the first two Magnéli phases Ti$_4$O$_7$ and Ti$_5$O$_9$ and corundum-phase Ti$_2$O$_3$.

\begin{center}
 \begin{figure}[t]
  \includegraphics[width=0.95\columnwidth]{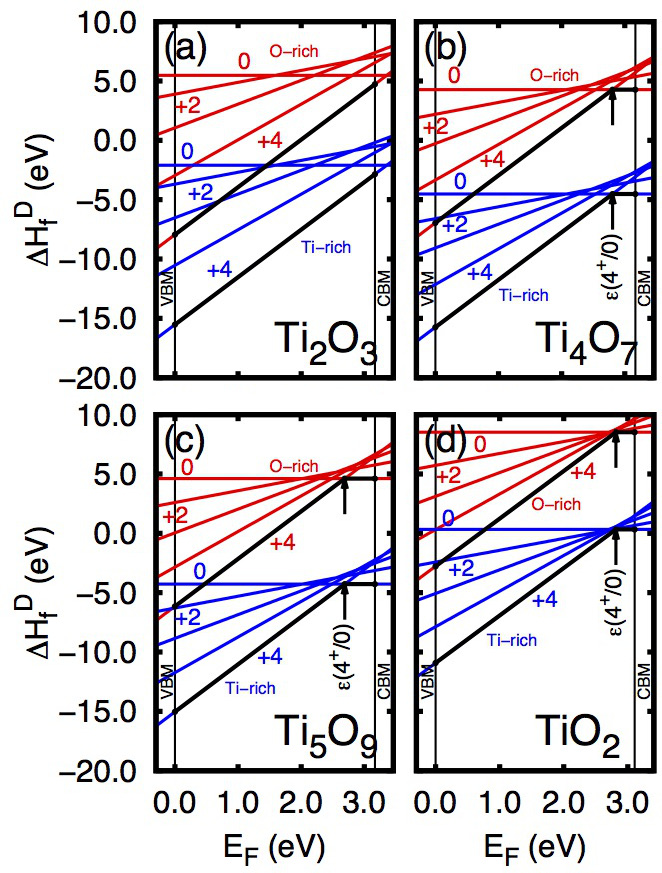}
  \caption{(color online) Formation enthalpies for (a) Ti$_2$O$_3$, (b) Ti$_4$O$_7$, (c) Ti$_5$O$_9$, (d) and $\text{Ti}_i$ point defect in TiO$_2$-rutile. The $\text{Ti}_i$ formation energies shown on (d) are from Ref. \cite{Lee2012}. The thick black lines emphasize the lowest energy charge state for each occupation for the entire band gap span, and the transitions from +4 charge state to the neutral charge state in Ti$_4$O$_7$, Ti$_5$O$_9$, and TiO$_2$ are also featured as $\varepsilon(4^+/0)$.}
    \label{fig:e-form}
 \end{figure}
\end{center}

The depletion of oxygen from TiO$_2$ and the ensuing formation of these oxygen-deficient phases can be described by two processes: (i) either the Magnéli phase is formed by the removal of oxygen and consequently the formation of ordered $V_{\text{O}}$ planes (secant to the $c$ vector, see figure \ref{fig:structs}) or (ii) from the formation of $\text{Ti}_i$ in ordered planes, \textit{i. e.},
\begin{align}
	n\text{TiO}_2 + V_{\text{O}} & \rightarrow \text{Ti}_n\text{O}_{2n-1} \label{eq:vacancy} \\
    (2n-1)\text{TiO}_2 + \text{Ti}_{\text{i}} & \rightarrow 2\text{Ti}_n\text{O}_{2n-1}. \label{eq:interstitial}
\end{align}

Our calculations include a reservoir of atoms at constant chemical potential $\mu_{\alpha}$ ($\alpha = \text{Ti}, \text{O}$) and a reservoir of electrons with the chemical potential at $E_F$. The formation enthalpy $\Delta H_f^D$ with respect to TiO$_2$ as a host material is given by
\begin{align}
	\label{eq:e-form-isolated}
	\Delta H_f^D (E_F, \mu, q) &= E_D(q) - E_H + \sum_{\alpha} m_{\alpha} \mu_{\alpha} + \nonumber \\
							   &+ q (E_F + E_{VBM} + \Delta V),
\end{align}
where $E_H$ and $E_D(q)$ are respectively the total energies of the system before (TiO$_2$) and after (Ti$_n$O$_{2n-1}$) exchanging $m_{\alpha}$ atoms with the reservoirs. The total energies $E_D$ and $E_H$ are obtained from density-functional calculations performed using the \texttt{VASP} code \cite{KresseFurthmuller1996} using the hybrid functional proposed by Heyd, Scuseria, and Ernzerhof (HSE)\cite{HSE}. The plane wave cutoff is set at 520 eV for all calculations and k-point sampling through the Brillouin zone was performed via the Monkhorst-Pack scheme. For charged systems, the unit cell volume is fixed as that of the uncharged system and atomic positions within the unit cells are relaxed. Our choice is justified by recent experimental results showing the formation of oxygen-deficient crystalline phases inside a TiO$_2$ matrix \cite{Kwon2010}. Test calculations where the unit cell was allowed to fully relax were performed, resulting in the same qualitative behavior (see supplementary material for further details). The Fermi energy $E_F$ is given with respect to the VBM of TiO$_2$ ($E_{VBM}$), and $\Delta V$ is a band-bottom alignment correction used to place all energies at the same reference, obtained from core level shifts \cite{Li2009}. For Ti atoms the $3p4s3d$ electrons were considered as valence electrons whereas the $2s2p$ configuration was considered for O atoms. Core level energies were obtained solving the Kohn-Sham equations for these inner level electrons subjected to a potential given by the pseudopotential method projector-augmented-wave (PAW) scheme \cite{Blochl1994}. In our calculations the chemical potential of oxygen was obtained from the $\text{O}_2$ molecule while the same quantity for the titanium atom was obtained from a bulk calculation of the hcp structure of metallic Ti.

Typically one uses equation \ref{eq:e-form-isolated} to address the formation enthalpy of defects. Here however we use this methodology to calculate the stability of the \textit{pseudo-defects} in Magnéli phases for different charge states $q$. The chemical potentials one should use for the expression could be either $\mu_{\text{O}}$ for the removed oxygen (Eq. \ref{eq:vacancy}; $m_{\text{O}} > 0$) or $\mu_{\text{Ti}}$ for the added titanium (Eq. \ref{eq:interstitial}; $m_{\text{Ti}} < 0$). We choose to discuss only the situation where the Magnéli and Corundum phases are formed via the insertion of Ti atoms (Eq. \ref{eq:interstitial}) because the electronic properties of the Ti$_n$O$_{2n-1}$ phases studied here exhibit \textit{pseudo-defect} properties as if the material were TiO$_2$-rutile doped by Ti interstitial. Moreover, the formation enthalpies for the reaction in Eq. \ref{eq:vacancy} can be obtained by using the oxygen chemical potential $\mu_{\text{O}}$, being the difference in the enthalpy curves in that case just constant shifts to the values presented here; the charge transfer properties remain identical.
\begin{center}
  \begin{figure*}[t]
  \includegraphics[width=0.95\textwidth]{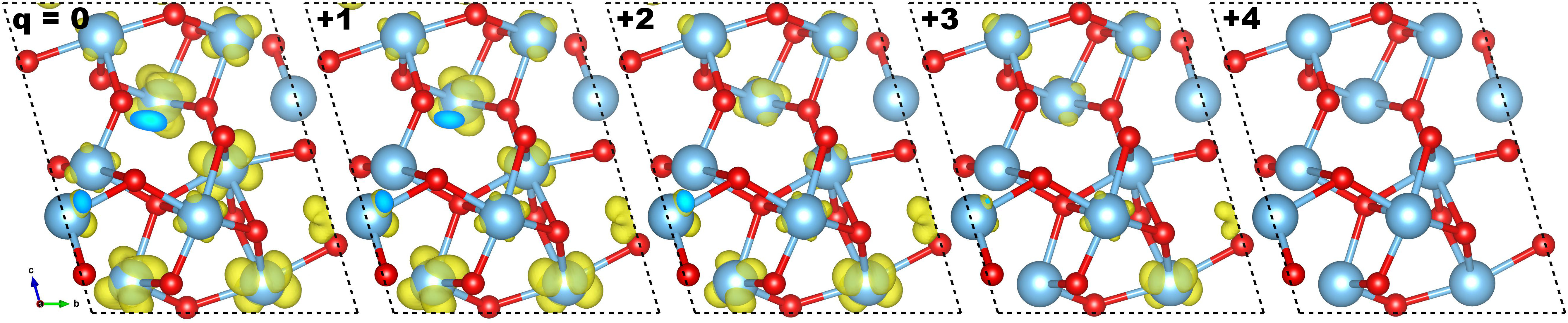}
   \caption{(color online) Real space projection of the \textit{intermediate band} on the PDOS of figure \ref{fig:dos-ti5o9}, from $E_F - 1.5$ eV to $E_F$ from all charged states with the exception of +4. The isosurfaces are depicted for all charge states, from the neutral case to +4 from left to right. We plot the same isosurface ($10^{-2}e \cdot \text{Bohr}^{-3}$). The +4 charge state presents no intermediate band, the structure is presented only for the sake of completion.}
   \label{fig:project}
  \end{figure*}
\end{center}

Formation enthalpies for the Ti$_2$O$_3$, Ti$_4$O$_7$ and Ti$_5$O$_9$ structures (when one considers $\alpha = \text{Ti}$ and $m_{\alpha} = +1$ in equation \ref{eq:e-form-isolated}), as well as the same data for the $\text{Ti}_i$ in TiO$_2$ rutile obtained from Lee \textit{et al.} \cite{Lee2012} are depicted in Figure \ref{fig:e-form}. O-rich and Ti-rich conditions are obtained by using the boundaries for $\mu_{\text{Ti}}$ given by the stability condition of each compound. Notice that Ti$_2$O$_3$ presents the full ionized charge state (+4) as the most stable (lowest formation energy) spanning the entire TiO$_2$-rutile band gap, while both Ti$_4$O$_7$ and Ti$_5$O$_9$ present the same trend from the VBM up to $\varepsilon(+4/0) = E_{\text{CBM}} - 0.36$ eV and 0.48 eV respectively. The position of $\varepsilon(+4/0)$ marks an abrupt transition from the +4 state to the neutral state. Interestingly this transition lies close to the very same $\varepsilon(+4/0)$ for the isolated $Ti_{\text{i}}$ in TiO$_2$-rutile (0.29 eV) \cite{Lee2012}. Thus our defect-free Ti$_n$O$_{2n-1}$ structures behave in a fashion similar to TiO$_2$ with intrinsic defects ($\text{Ti}_i$ or $V_{\text{O}}$). An important distinction must be taken at this point, since the presence of extrinsic defects can lead to different behaviors. For example, nitrogen impurities in the Magnéli phases lead to an electron-hole compensation effect that can significantly alter properties such as the bandgap and photocatalytic activity \cite{Niu2015}. We emphasize the fact that we vary the occupation of the \textit{pseudo-defect} induced \textit{intermediate band} in a similar fashion as it is done to mid gap states of isolated defects inside semiconductors. Moreover both Ti$_4$O$_7$ and Ti$_5$O$_9$ present negative-$U$ behavior \cite{Watkins1984}. This is also the case of $\text{Ti}_i$ point defects inside TiO$_2$-rutile \cite{Lee2012}. 

The nature of the charged state in the Ti$_n$O$_{2n-1}$ structures can be understood from the Projected Density of States (PDOS) and real-space projections of selected states. Figures \ref{fig:dos-ti5o9} and \ref{fig:project} show this kind of analysis for Ti$_5$O$_9$ as a point in case---Ti$_4$O$_7$ presented a similar behavior (see supplementary material). The neutral structure shows a midgap \textit{intermediate band} akin to isolated defect states. These states are mostly of Ti(d) character---as are the unoccupied bands---delocalized over several Ti atoms, as shown in Figure \ref{fig:project}. It is known from literature that 3d transition metal related defects exhibit multiple charged states \cite{Haldane1976,Raebiger2014} as is the case of $Ti_{\text{i}}$ in TiO$_2$-rutile \cite{Lee2012}. Recently, such charge transitions have also been observed for extended defects \cite{Raebiger2014}. Here, we show that even perfect crystals that deviate from stoichiometry may exhibit similar charge states. The d orbital rehybridization seen in Figure \ref{fig:project} suggest that these multiple charge states of the \textit{pseudo-defects} in the Magnéli phases are facilitated by a self-regulating response mechanism \cite{Raebiger2008,Wolverton1998,Haldane1976} which also explains why the material does not undergo a Coulomb catastrophe.

To estimate the storage capacity of these Magnéli phases, we consider a maximum of 4 holes per \textit{pseudo-defect} corresponding to the quadruple donor transition observed, as well as the maximum capacity of the intermediate band to accommodate 4 electrons---two electrons for each of the two $V_{\text{O}}$'s, or alternatively, four electrons for a single $Ti_{\text{i}}$, according to the previous discussion. Using this and considering a device operating at a 1 V potential, the theoretical maximum capacitance is approximately 1300 F/g for Ti$_2$O$_3$, 600 F/g for Ti$_4$O$_7$, and 500 F/g for Ti$_5$O$_9$, placing those systems at par with materials used to build supercapacitors. As discussed earlier, by interfacing these oxygen deficient phases appropriate leads, one can control its charge state.

In conclusion, we have performed electronic structure DFT calculations to asses the formation and electric charging of the TiO Magnéli and corundum phases. We show that these materials contain \textit{pseudo-defects}, \textit{i.\ e.}, they behave akin to Ti$_{\text{i}}$ doped TiO$_2$-rutile with a concentration of $\nicefrac{1}{2}$ quadruple donor defects per formula unit Ti$_n$O$_{2n-1}$. These \textit{pseudo-defects} are characterized by an \textit{intermediate band} that can be charged, thus, the material can become charged and used for high-capacity charge storage. We propose that the same behavior shown here for the oxygen deficient TiO phases exists in other semiconductor materials. The required condition is the presence of the \textit{intermediate band} with a large enough density of states, which we expect to be the case in other materials that present stable phases over a wide range of stoichiometries.

\begin{acknowledgments}
This work was supported by FAPESP and CNPq. The support given by Cenapad-SP in the form of computational infrastructure is also acknowledged.
\end{acknowledgments}

%

\end{document}